# Hardware Accelerators for Artificial Intelligence

S M Mojahidul Ahsan, Anurag Dhungel, Mrittika Chowdhury, Md Sakib Hasan,
Tamzidul Hoque

Email: ahsan@ku.edu

In this chapter, we explore the specialized hardware accelerators designed to enhance
Artificial Intelligence (AI) applications, focusing on their necessity, development,
and impact on the AI field. It covers the transition from traditional computing
systems to advanced AI-specific hardware, addressing the growing demands of
AI algorithms and the inefficiencies of conventional architectures. The discussion
extends to various types of accelerators, including GPUs, FPGAs, and ASICs, and
their roles in optimizing AI workloads. Additionally, it touches on the challenges and
considerations in designing and implementing these accelerators, along with future
prospects in the evolution of AI hardware. This comprehensive overview aims to
provide readers with a clear understanding of the current state and future trends in
AI hardware development, ensuring it is accessible to both experts and beginners in
the field.

## 1.1 Introduction to Hardware Accelerators for AI

### 1.1.1 Overview of AI Advancements and Impacts

Artificial intelligence (AI) and deep learning have emerged as powerful forces driving
innovation across a vast spectrum of industries and applications. From revolution-
izing the way we interact with technology to tackling complex scientific challenges,
these advancements are shaping the future of our world. Fig. 1.1 shows the transfor-
mation of AI over the past decade marked by a series of groundbreaking innovations,
beginning with the introduction of AlexNet in 2012, which revolutionized the field
of deep learning. This journey has seen the development of increasingly sophis-
ticated algorithms, culminating in the latest generation of large language models
(LLMs). These advancements have significantly expanded AI capabilities, enabling
more complex and nuanced understanding and generation of human language, and
transforming a wide range of applications across industries.

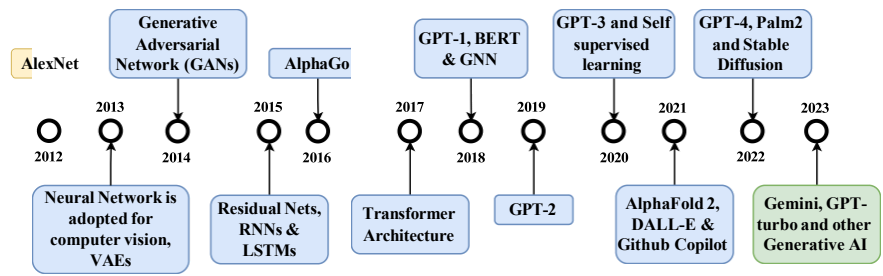

**Fig. 1.1**  The evolution of Artificial Intelligence over the past decade, marking significant milestones from the introduction of AlexNet in 2012, which revolutionized neural networks, to the development of sophisticated Generative Pre-trained Transformer (GPT) models.

In the current landscape, several key factors stand out as crucial for the ongoing advancement of AI.

- **Availability of data:** The explosion of data across various fields has provided fertile ground for the training of deep learning models. From social media posts and medical records to satellite imagery and financial transactions, the abundance of data has been instrumental in fueling progress in AI.
- **Increased computational power:** Advancements in hardware, such as Graphics Processing Units (GPUs) and specialized AI accelerator chips, have greatly enhanced the building and training of increasingly sophisticated neural networks. This increase in processing capability enables faster training durations and the handling of larger datasets, continuously expanding the horizons of AI potential.
- **Deep learning architectures:** Deep neural networks, including convolutional neural networks (CNNs) and recurrent neural networks (RNNs), have excelled in various applications such as image classification, natural language understanding, and language translation. Their success in these domains has significantly advanced the capabilities of AI in processing and interpreting visual and textual data.
- **Generative AI:** This subfield of AI focuses on creating new data, such as text, images, code, and music. Generative models like Generative Pre-trained Transformers (GPTs) can generate human-quality content that is often indistinguishable from the real thing. This has opened up exciting possibilities for creative applications, personalized content generation, and data augmentation.

## 1.1.2  AI Hardware Accelerators: Overcoming Traditional Limits

The world of artificial intelligence is rapidly evolving, demanding ever-increasing computational power to handle complex algorithms and massive datasets. Traditional computing architectures, such as the Von Neumann architecture, have long served as the backbone of information processing. However, these architectures are increasingly falling short when it comes to the specific demands of modern AI, particularly in the realm of neural network processing.

- **Bottlenecks of Traditional CPUs**
  Traditional Central Processing Units (CPUs) with Von Neumann architecture face significant challenges with deep learning, primarily due to inefficiencies in managing the high volume of multiply-accumulate (MAC) operations essential for neural network computation. This issue is compounded by the limited parallel processing capabilities, narrower vector processing units, and lower memory bandwidth of modern CPU architectures creating bottlenecks in data handling. Additionally, the complex cache hierarchies introduce latency, and the general-purpose design of CPUs, including their instruction sets and control flows, leads to underutilization in the predictable, repetitive tasks of neural network processing [15]. These factors, combined with the energy-intensive nature of neural network computations, underscore the inadequacy of traditional computing systems for the parallel, data-heavy demands of AI workloads, especially as the depth and complexity of neural networks continue to grow.

- **GPUs: A Stepping Stone, but Not the Solution**
  GPUs, with their thousands of cores, excel at parallel processing, allowing them to perform a large number of simple computations simultaneously, which is a common requirement in AI and deep learning algorithms. Moreover, GPUs incorporate cutting-edge memory solutions, such as High Bandwidth Memory (HBM) and GDDR6X, to support the intensive data transfer required for these computations. This high bandwidth enables GPUs to efficiently perform parallel processing tasks, such as matrix multiplications, and accumulate operations found in deep learning algorithms, thereby significantly speeding up AI model training and inference processes. Despite their superiority over CPUs in AI applications, GPUs are increasingly overshadowed by their energy-intensive nature as AI algorithms grow in complexity and data requirements. This substantial power consumption stems from their adherence to the Von Neumann architecture, which leads to notable limitations in power efficiency, memory bandwidth, and latency. These constraints not only escalate operational costs but also impede the scalability and sustainability of AI advancements [16]. Consequently, while GPUs represent a significant advancement over CPUs, they are not the ultimate solution for the evolving landscape of AI, particularly in an era where energy efficiency and area benchmarks are paramount.

- **Need for Specialized Hardware Accelerators: Tailored for AI Efficiency**
  Recognizing the shortcomings of traditional architectures, researchers have turned to the development of specialized hardware accelerators for AI. These accelerators are intended to address the unique computational requirements of AI in their design, focusing on the following key aspects:

  - **High Parallelism:** AI accelerators typically employ massive parallelism, utilizing hundreds or even thousands of processing cores to handle the massive volume of MAC operations simultaneously. This dramatically increases efficiency and performance compared to the conventional computing methods.

– **Near-Memory Processing:** To address the bottleneck of data movement between external memory and the processing unit, AI accelerators employ near-memory or in-memory computing. This approach significantly reduces latency and energy consumption by minimizing the distance data travels, and processing data closer to where it is stored.

– **High Memory Bandwidth:** Essential for supporting the high-throughput demands of AI workloads, specialized accelerators are equipped with high memory bandwidth. This feature enables the rapid transfer of data between memory and processing units, crucial for feeding the computational appetite of AI algorithms and preventing processing cores from becoming data-starved.

– **Energy Efficiency:** At the heart of AI accelerator design, achieving energy efficiency is paramount. This is accomplished through a multi-layered optimization strategy, spanning from refined parallel algorithms and circuit-level enhancements in memory and computational units to cutting-edge device-level breakthroughs, such as the integration of energy-saving non-volatile memory technologies for synaptic functions.

– **Reduced Instruction Set Computing (RISC):** Unlike general-purpose CPU architectures, AI accelerators often utilize RISC architectures. These architectures are designed for simpler, dedicated instructions, resulting in faster execution and reduced energy consumption. Moreover, AI accelerators may use simplified or specialized instruction sets that are optimized for computations most frequently performed in AI algorithms.

– **Reduced Precision Arithmetic:** AI applications often exhibit inherent resilience to noise and imperfection. Leveraging this characteristic, AI accelerators frequently employ reduced precision arithmetic, sacrificing some accuracy for significant gains in computational efficiency and memory bandwidth. The specialized AI hardware is capable of using this technique for efficient computing.

The shift toward specialized hardware for AI acceleration is driven by the limitations of traditional computing architectures in keeping pace with the demands of advanced AI and neural network algorithms. This move is critical for harnessing the full potential of AI, enabling more efficient and powerful processing capabilities tailored for the complexity and depth of modern AI computations. Such dedicated hardware is essential for the next wave of AI innovations, providing the foundation for more sophisticated, real-time AI applications and significantly advancing the field.

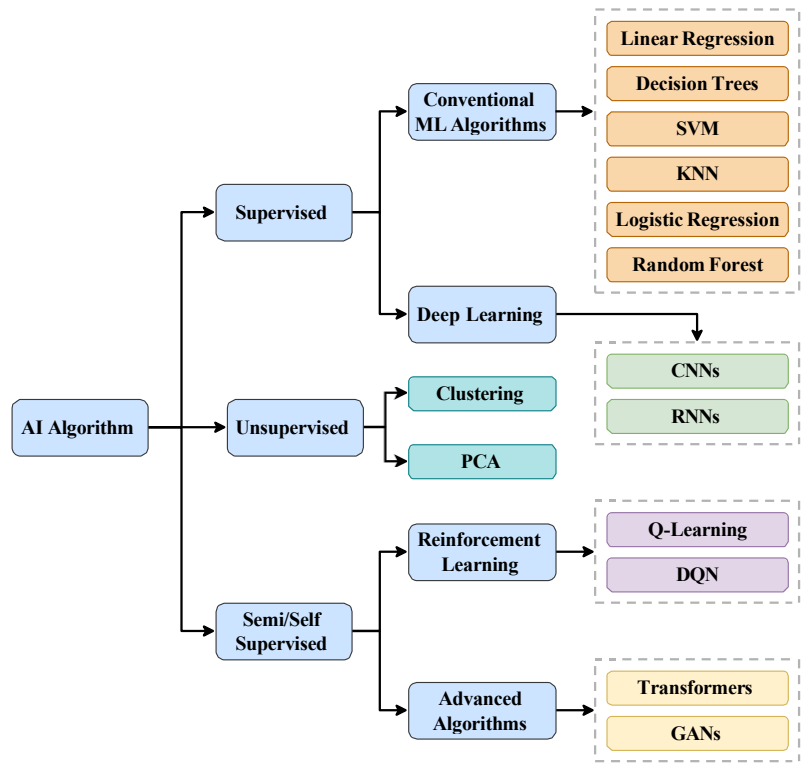

**Fig. 1.2** A classification of key AI Algorithms.

## 1.2 AI Algorithms and their Hardware Implementation

### 1.2.1 Overview of key AI algorithms:

AI algorithms, serving as the cornerstone of contemporary technological break-throughs, are designed to tackle a diverse array of tasks ranging from intricate data analysis to the autonomous operation of systems. These sophisticated algorithms demand extensive computational power to function effectively, highlighting the pivotal role of computational resources in enabling their full potential. The synergy between advanced AI algorithms and the computational infrastructure paves the way for groundbreaking developments in AI, fostering the creation of more efficient, powerful, and adaptable applications across various fields. The interplay between these sophisticated algorithms and their hardware implementation is crucial in advancing the field of AI, leading to more efficient, powerful, and versatile applications. A classification of modern key AI algorithms are shown in Fig. 1.2. Let's dig deeper into the functionalities and applications of key AI algorithms, categorized into three major areas:

## 1  Supervised Learning

• **Conventional Machine Learning:**

- **Linear Regression:** This algorithm learns a linear relationship between features and a continuous target variable. It operates by fitting a linear equation to the data points, optimizing the line to minimize the sum of the squared differences between the predicted and actual values. This straightforward yet effective method is widely used in regression tasks such as predicting housing prices based on factors like size and location.
- **Decision Trees:** This algorithm classifies data points by asking a series of yes/no questions based on features. Each question splits the data into smaller subsets until reaching a final classification. Decision trees offer interpretable results and are efficient for handling large datasets.
- **Support Vector Machine (SVM):** SVM is a powerful classification algorithm that finds the best hyperplane separating different classes in the feature space. It maximizes the margin between data points of different categories, offering robustness in classification tasks.
- **K-Nearest Neighbors (KNN):** The K-Nearest Neighbors (KNN) algorithm classifies a data point based on the majority vote of its 'k' nearest neighbors in the feature space. Versatile and straightforward, KNN is used for both classification and regression tasks by analyzing the distances between data points to find the most similar neighbors. Its simplicity and effectiveness make it popular in various machine learning applications.
- **Logistic Regression:** Logistic regression is used for binary classification. It estimates the probability that a given input point belongs to a certain class. This is done by applying a logistic function to a linear equation, providing a probabilistic foundation for classification.
- **Random Forest:** Random Forest is an ensemble learning technique that builds multiple decision trees and combines their results for improved accuracy and stability. It is effective for both classification and regression tasks, reducing overfitting and enhancing predictive performance.

• **Deep Learning:**

- **Convolutional Neural Networks (CNNs):** These networks excel at image recognition and computer vision tasks. They learn to extract features from images by applying convolution filters over the input data. CNNs are widely used in applications like face recognition, object detection, and medical image analysis.
- **Recurrent Neural Networks (RNNs):** These networks are developed to process sequential data like text and speech. They have internal loops that allow them to remember information from previous inputs, making them ideal for tasks like language translation and machine translation.

## 2  Unsupervised Learning

– **Clustering:** Clustering algorithms group similar data points based on their characteristics, identifying patterns and structures in unlabeled data. This technique is useful for tasks such as segmenting customers by purchase history. Common clustering methods include K-means and hierarchical clustering, which organize data into meaningful groups to uncover insights and trends.

– **Principal Component Analysis (PCA):** This algorithm reduces the dimensionality of data by identifying the most important features. By projecting the data onto a lower-dimensional space, PCA helps visualize data and improve the performance of other algorithms.

## 3 Semi/Self Supervised Learning

• **Reinforcement Learning:**

– **Q-Learning:** This algorithm learns an optimal policy for an agent in a given environment. It estimates the expected reward for taking a specific action in each state and chooses the action with the highest expected reward. Q-Learning has been successfully applied in robotics and game playing.

– **Deep Q-Networks (DQN):** This algorithm utilizes Q-Learning with a neural network to handle complex state spaces. DQN has achieved superhuman performance in various Atari games, demonstrating its potential for solving complex decision-making problems.

• **Advanced Algorithms:**

– **Transformers:** Transformers are advanced neural networks that utilize the attention mechanism to evaluate relationships within an input sequence. This capability makes them highly effective for natural language understanding and generation tasks. They are integral to large language models like BERT and GPT-3, which excel in applications ranging from text translation to conversational agents.

– **Generative Adversarial Networks (GANs):** GANs feature a generator that creates new data and a discriminator that evaluates its authenticity. This adversarial process enhances the performance of both networks, leading to the creation of highly realistic outputs such as images, music, and text. By constantly challenging each other, the generator produces increasingly convincing data, while the discriminator becomes more adept at identifying fakes, resulting in the progressive improvement of the system's capabilities.

These algorithms' computational complexity and requirements are a driving force behind the evolution of specialized hardware. The hardware not only needs to handle vast amounts of data efficiently but also provides the speed necessary for real-time processing and learning.

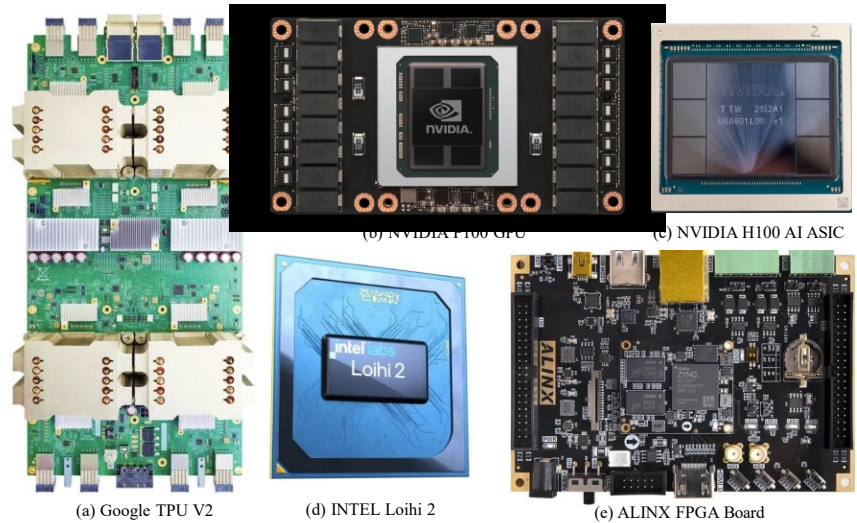

**Fig. 1.3** Modern Hardware Accelerators: (a) TPU v2 was unveiled at Google I/O in May 2017 [1], (b) NVIDIA Tesla P100: The World's First AI Supercomputing Data Center GPU [2], (c) Nvidia H100 Hopper chip [3], (d) Intel Loihi 2 Neuromorphic Chip [4], (e) ALINX AX7Z020: SoC FPGA Development Board [5]

### 1.2.2  Case Studies of Hardware Accelerator for AI

Artificial intelligence (AI) algorithms are increasingly being deployed in various domains, such as computer vision, natural language processing, robotics, and healthcare. However, the computational complexity and resource requirements of these algorithms pose significant challenges for their efficient and scalable implementation on hardware platforms. In this section, we review some case studies of specific hardware specially designed for handling fast and energy-efficient AI computation, covering different types of AI algorithms and hardware platforms. Selected hardware packages and boards are demonstrated in Fig. 1.3.

- **Tensor Processing Unit (TPU):** One example of a state-of-the-art AI algorithm implementation on a hardware accelerator is the Google TPU (Tensor Processing Unit), a custom ASIC (Application-Specific Integrated Circuit) specifically designed to accelerate deep neural networks (DNNs). The TPU uses a systolic array architecture, facilitating high-throughput matrix multiplication, essential for DNN operations. It also supports reduced-precision arithmetic and data compression techniques, enhancing both the performance and energy efficiency of DNN inference. The TPU has been deployed in powering various Google services, including Google Search, Google Photos, Google Translate, and AlphaGo. A detailed architectural explanation of the TPU is provided in a later section1.3.

- **IBM TrueNorth and NorthPole:** Another example of a neuromorphic chip specifically designed for faster and energy-efficient AI computation is the IBM TrueNorth chip. This neuromorphic processor emulates the structure and function of biological neurons and synapses. TrueNorth [10] contains 4096 cores, each with 256 neurons and a 256×256 binary synapse array. It operates asynchronously, generating and propagating spikes among neurons. TrueNorth supports neural network models like CNNs, RNNs, and SNNs, and is applied in image recognition, video analysis, and anomaly detection. A recent advancement is IBM's NorthPole chip [11]. It features a neural-inspired architecture for enhanced performance, energy, and area efficiency compared to comparable architectures. NorthPole integrates computation with memory, reducing off-chip memory requirements. It's a digital system allowing for tailored bit precision, optimizing power usage. This chip, using a 12-nanometer process, achieves significant improvements in energy, space, and time metrics over prevalent architectures, even those using more advanced technology processes.

- **Intel Loihi Series:** The Intel Loihi chip, a neuromorphic processor, mirrors the behavior and learning processes of biological neurons and synapses. It features 128 cores, each with 1024 neurons and 128K synapses, programmable for various tasks. The chip is adept at adapting to new inputs and environments, thanks to its support for spike-timing-dependent plasticity and online learning. Loihi is compatible with multiple neural network architectures like CNNs, RNNs, and SNNs. Its applications include gesture recognition, path planning, reinforcement learning, and optimization. The Intel Loihi 2 neuromorphic chip is a significant evolution from its predecessor, offering up to 10 times the performance. It supports more advanced and flexible neuromorphic computing architectures, allowing for a wide range of applications. Loihi 2 introduces improved features such as generalized event-based messaging, enhanced neuron model programmability, and advanced learning capabilities. These upgrades enable the chip to support more complex algorithms and applications, positioning it at the forefront of neuromorphic computing research and development.

- **Graphics Processing Unit (GPU):** In addition to specialized hardware, advanced GPUs are crucial in AI advancements, particularly in data centers. NVIDIA's P, H, and V series GPUs, for example, excel in accelerating DNNs, especially during training phases. Their parallel architecture allows for numerous simultaneous floating-point operations, ideal for DNN processes. GPUs also feature high memory bandwidth and performant interconnects, handling vast data quantities efficiently. They can be combined with CPUs or FPGAs, creating heterogeneous systems that leverage the strengths of various hardware components.

- **Field-Programmable Gate Arrays (FPGAs):** FPGAs are increasingly recognized for their critical role in accelerating deep learning algorithms. Unlike traditional hardware, FPGAs offer a unique blend of programmability and performance, making them highly adaptable to the evolving needs of AI computations. Their ability to be reconfigured for specific tasks allows for optimized processing of complex neural networks, enhancing both efficiency and speed. This adaptability, coupled with their capacity for parallel processing, positions FPGAs as a pivotal technology in advancing the field of AI, particularly in areas requiring real-time processing and energy-efficient solutions.

- **Memory-Centric Computation:** In-memory computing, leveraging traditional memory technologies such as SRAM and DRAM, represents a promising approach to addressing the von Neumann bottleneck. This paradigm integrates computation directly within the memory units, reducing data transfer latency and energy consumption. For instance, SRAM-based in-memory computing designs enable fast and energy-efficient operations by conducting logical and arithmetic functions directly within the memory arrays [32]. Similarly, DRAM-based architectures utilize the inherent parallelism and large storage capacities of DRAM to accelerate AI workloads [33]. These memory-centric approaches can significantly enhance the performance of AI applications by minimizing the data movement between the CPU and memory, thus overcoming one of the critical limitations of conventional von Neumann architecture.

- **Emerging NVM-based Process-in-memory:** Emerging memory technologies like RRAM, PCM, and MRAM enable a new paradigm called process-in-memory (PIM). This architecture integrates memory and processing elements within the same chip, enabling data to be processed directly within the memory, and eliminating the need for data movement between memory and processing units. This drastically reduces energy consumption and improves performance, making PIM a key technology for future AI hardware.

  - **Resistive Random Access Memory (RRAM):**RRAM utilizes the resistance change of materials to store data, offering high density and low power consumption. This makes it promising for resource-constrained edge AI applications. For example, RRAM-based accelerators have been developed for image recognition tasks on mobile devices, achieving significant performance improvements compared to traditional CPUs and GPUs [12].

  - **Phase Change Memory (PCM):** PCM stores data by reversibly switching a material's phase between crystalline and amorphous states. This allows for high performance, low power consumption, and non-volatility, making it suitable for various AI applications. PCM-based accelerators have demonstrated promising results in tasks like natural language processing and deep learning, offering faster processing and reduced energy consumption compared to traditional memory technologies [13].

  - **Magnetoresistive Random-access Memory (MRAM):** MRAM uses magnetic fields to store data, offering low latency, energy efficiency, and non-volatility [14]. This makes it ideal for applications requiring fast and reliable data access, such as real-time AI applications. MRAM-based accelerators are currently developing, with potential applications in areas like autonomous vehicles and robotics, where fast and accurate decision-making is critical.

  - **Ferroelectric Field-Effect Transistor (FeFET):** FeFET leverages the polarization properties of ferroelectric materials to store data, combining non-volatility with high speed and low energy consumption [14]. This unique combination makes FeFET particularly appealing for AI applications that demand efficient data processing and storage capabilities. FeFET-based accelerators are emerging as a potent solution for AI tasks, offering significant advantages in terms of density, power efficiency, and speed. Early implementations of FeFET technology in AI hardware have shown promising results.

These emerging devices create new research opportunities for PIM technology. Tools like NVSim [29], NeuroSIM [30], and NVMexplorer [31] can be utilized to investigate these devices for various innovative applications.

To summarize, a variety of creative solutions are being employed to address the computational needs of AI applications in different areas. It covers a range of innovations, from Google's TPU with its systolic array design that optimizes deep neural network tasks to IBM's brain-inspired neuromorphic chips. Each example illustrates a distinct method for boosting both efficiency and energy conservation. Intel's Loihi chips and high-end GPUs merge neuromorphic concepts with parallel computation, whereas FPGAs provide versatility and adaptiveness. Furthermore, the advent of NVM-based technologies marks the beginning of a new phase in memory-centric computing designs, setting the stage for substantial progress in AI hardware. The next section will provide a side-by-side comparison of these technologies, evaluating their respective strengths and weaknesses.

### 1.2.3 Comparative Analysis of Different Hardware Solutions for AI:

The rapid evolution of artificial intelligence (AI) has spurred an equally significant demand for specialized hardware solutions capable of handling its increasingly sophisticated algorithms. While traditional CPUs have played a crucial role in AI development, they are often outmatched by the computational demands of modern machine learning (ML) and neural network (NN) algorithms. This has led to the rise of alternative hardware platforms specifically designed to accelerate AI tasks. Here, we will explore these diverse hardware solutions, providing insight into their comparative strengths and weaknesses.

**A. GPUs:**

– **Strengths:** Initially designed for rendering graphics in gaming and visual applications, GPUs have evolved significantly over the past 30 years to become crucial for AI applications. The high data throughput and massive parallelism of GPUs, with their hundreds of cores, allow them to excel in parallel calculations such as matrix multiplications. This makes them ideal for deep learning, big data analytics, and genomic sequencing. In particular, GPUs are highly effective in training AI models, where the ability to process large, similar datasets simultaneously is essential. Extensively used in neural networks and accelerated AI operations, GPUs provide the computational power necessary for handling large volumes of identical or unstructured data. Today, their ad-vanced capabilities make them a staple in data centers and cloud applications, continuing to play a significant role in the AI revolution.

– **Weaknesses:** GPUs, while powerful, face challenges in multi-GPU setups due to complexity and scalability issues. Programming for these systems requires specialized knowledge, and communication overhead between GPUs can hinder performance. Memory management is critical, as large datasets may not fit in a single GPU. Additionally, the high cost and increased power consumption are significant considerations. Compatibility with AI frameworks and diminishing performance gains with added GPUs further complicate their use. Maintenance and limited availability of cloud services also pose challenges.

## B. FPGAs (Field-Programmable Gate Arrays):

– **Strengths:** FPGAs, positioned between CPUs and GPUs, offer a unique blend of benefits for AI acceleration. Unlike CPUs with limited parallelism and GPUs that may lack power efficiency, FPGAs provide reconfigurable hardware tailored to specific AI tasks. They enable custom hardware accelerators that can be programmed to align perfectly with an AI algorithm's requirements, resulting in enhanced performance and energy efficiency. FPGAs excel due to their parallel processing capabilities, ideal for AI operations like matrix multiplications. Additionally, they allow for bespoke hardware customization, offloading intensive computations for faster execution. Crucially, FPGAs are well-suited for low-latency inference in real-time applications, offering immediate predictions. Their ability to be fine-tuned for specific tasks also makes them highly energy-efficient, particularly beneficial for edge computing and IoT applications.

– **Weaknesses:** Despite their versatility, FPGAs pose challenges in programming complexity, requiring expertise in hardware description languages like VHDL or Verilog, which can be a barrier for some AI practitioners. They also face resource constraints, including limited logic gates, memory, and DSP blocks, making optimization of AI algorithms within these confines a complex task. Moreover, scalability with FPGAs in large clusters or cloud environments can be more intricate compared to GPUs. FPGAs are highly efficient for specific algorithms or tasks, but adapting them to different AI models may necessitate substantial reconfiguration, limiting their flexibility in certain scenarios.

## E. Neuromorphic Integrated Circuits (ICs):

– **Strengths:** Neuromorphic ICs, inspired by the human brain, excel in efficiency and speed for specific AI tasks. Their design, mimicking neurons and synapses, allows for lower power consumption and faster data processing, particularly in pattern recognition and sensory data interpretation. This makes them ideal for edge computing applications in AI.

– **Weaknesses:** The complexity of mimicking biological structures means neuromorphic ICs are still in the early development stages. They may not yet match the versatility or raw power of more established platforms like GPUs for a broad range of AI tasks. Their specialized nature also poses a challenge for programming and integration into existing technology stacks.

**C. Application-Specific Integrated Circuits (ASICs):**

– **Strengths:** AI ASICs are custom hardware designed and optimized for specific AI computations, allowing for the creation of innovative architectures tailored to particular models or applications. This specialization in task execution leads to exceptional energy efficiency and a smaller footprint compared to other accelerators. Their design focus on specific functions also results in higher speed and performance, making ASICs highly effective for high-performance, energy-sensitive applications within AI. The combination of tailored architecture, power efficiency, and compact size positions ASICs as a superior choice for specialized AI computations.

– **Weaknesses:** The biggest drawback of ASICs is their lack of flexibility. Their custom design locks them into performing a single function, making them unsuitable for adapting to evolving AI algorithms and changing requirements. Additionally, their high development cost can make them less attractive for smaller projects.

**D. Emerging Devices:**

– **Strengths:** Emerging devices in AI hardware applications are advancing with novel architectures and memory technologies. Process-in-memory and near-memory computing, for instance, are changing how data is handled, bringing computation closer to storage, thereby improving speed and efficiency. Innovations in non-volatile memory technologies like ReRAM, PCM, and MRAM are also pivotal, as they enable matrix-vector multiplication operations directly within the device, a critical operation in AI algorithms. Additionally, new devices like FeFET (Ferroelectric Field Effect Transistor) are emerging, promising advancements in AI hardware design due to their unique properties. These technologies and architectures collectively represent a significant leap forward in AI hardware, offering improvements in speed, energy efficiency, and computing power.

– **Weaknesses:** Emerging device-based AI hardware accelerators, despite their potential, face challenges like variation, aging, and device variabilities, leading to inconsistent behavior. Process compatibility issues also present hurdles in integrating these technologies into existing manufacturing systems. However, engineers and physicists are actively working to optimize these devices for AI hardware design. They are focused on addressing these variabilities and compatibility issues to harness the full potential of these advanced technologies for efficient and reliable AI computation.

The diverse landscape of hardware solutions for AI presents developers and researchers with a range of options. Understanding the strengths and weaknesses of each platform and carefully considering application requirements is key to selecting the optimal hardware for successful AI implementation. As the field of AI contin-ues to evolve, we can expect further innovations in hardware design, pushing the boundaries of performance and efficiency to new heights.

## 1.3 AI Hardware Accelerator Architectures

In this section, we will discuss some of the significant advancements in chronological order, exploring the innovative landscape of AI accelerator architectures developed by research communities. These architectures have significantly contributed to the evolution of specialized AI processing hardware. Tailored specifically for the demands of AI and machine learning algorithms, they represent a departure from traditional computing paradigms, focusing instead on optimizing computational efficiency, reducing power consumption, and accelerating AI workloads. From the integration of novel memory technologies to the adoption of specialized computational models like neuromorphic computing and processing-in-memory (PIM), these architectures showcase the forefront of hardware design tailored to the unique requirements of AI applications. This chronological examination highlights the progressive advancements and groundbreaking contributions that have shaped the current state and future directions of AI accelerator architectures.

### 1.3.1 NeuFlow Architecture

The NeuFlow architecture, introduced in 2011 [18], is a dynamically reconfigurable dataflow processor tailored for vision-related tasks and implemented on FPGA. It features a scalable hardware structure and employs a dataflow compiler called luaFlow to convert high-level algorithmic flow-graphs into machine code. NeuFlow is engineered to handle real-time object detection, classification, and localization in intricate scenes, providing notable improvements in processing speed and energy efficiency.The architecture uses a 2D grid of Processing Tiles (PTs), each with processing operators and routing multiplexers, and a Smart DMA for efficient off-chip memory interfacing. It achieves high throughput by exploiting parallelism within modules and across images, with a focus on filter-based algorithms like convolutional networks. The NeuFlow system demonstrates a new approach to dataflow computing in vision systems, combining high performance with low power consumption. An illustration of this architecture is shown in Fig. 1.4. The NeuFlow architecture operates as follows:

1  High-Level Programming: NeuFlow begins with a high-level programming interface, where algorithms are represented as flow-graphs.
2  LuaFlow Compiler: These flow-graphs are translated into machine code by the LuaFlow compiler.
3  Processing Tiles (PTs): The machine code is executed on a 2D grid of PTs, each consisting of processing elements and routing multiplexers.

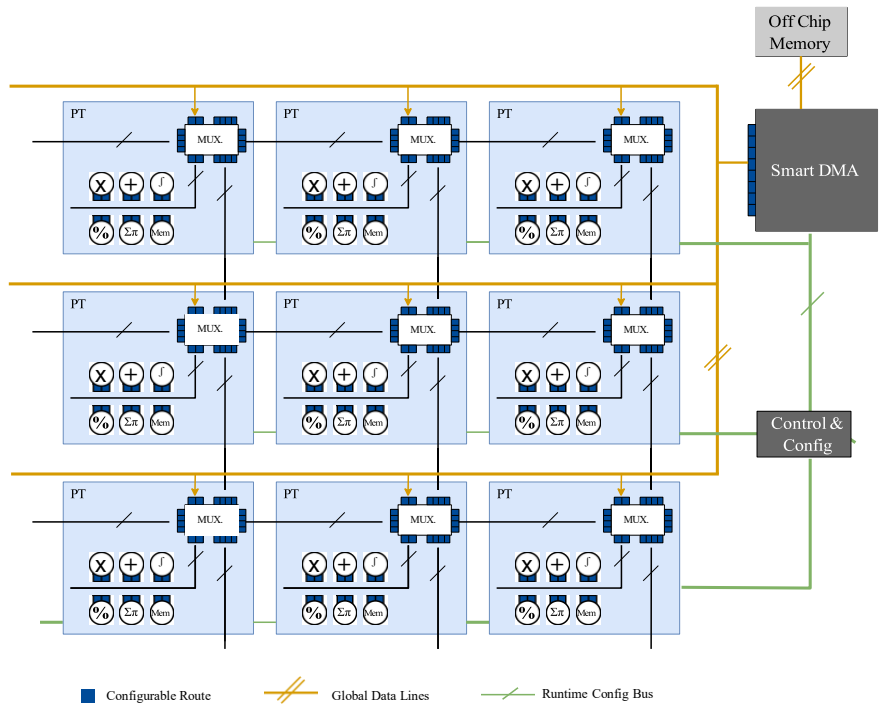

**Fig. 1.4** Dynamic Dataflow Architecture for NeuFlow (the figure is recreated from [18]).

4 Parallel Processing: NeuFlow exploits parallelism both within modules and across images for efficient processing.
5 Smart DMA: This component handles off-chip memory interactions, ensuring efficient data transfer and storage.
6 Real-Time Execution: The architecture is optimized for real-time vision tasks, providing high throughput for filter-based algorithms like convolutional networks.

NeuFlow's design allows for efficient, real-time processing of complex vision algorithms, highlighting its potential in advanced vision systems.

### 1.3.2 The DianNao Series:

The DianNao family of hardware accelerators, which falls under the ASIC category, is designed for machine learning, especially neural networks, and includes DianNao, DaDianNao[20], ShiDianNao[21], and PuDianNao[22]. Each architecture is specialized for specific aspects of machine learning, with a focus on minimizing memory transfers and maximizing efficiency. DianNao, the first in the series, is optimized for neural network computations. DaDianNao is a larger-scale system for handling more extensive neural networks. ShiDianNao focuses on low-power, embedded system applications, particularly for image processing. PuDianNao extends the capabilities to a broader range of machine learning techniques beyond neural networks. These architectures demonstrate significant improvements in speed and energy efficiency over traditional computing systems, notably in handling neural network computations. The baseline DianNao architecture is centered around its Neural Functional Units (NFUs), which are crucial for performing arithmetic operations integral to neural networks. These NFUs are designed to efficiently carry out multiplications and additions, operations that are fundamental in neural network computations. To support this processing, DianNao incorporates specific data buffers: Input, Weight, and Output Buffers. These buffers play a key role in reducing the dependency on external memory access, thereby enhancing overall computational efficiency. The data in DianNao flows through these NFUs in a pipelined manner. This pipeline approach allows for the simultaneous processing of multiple data points, significantly speeding up the neural network computations and ensuring a more efficient flow of data through the system.

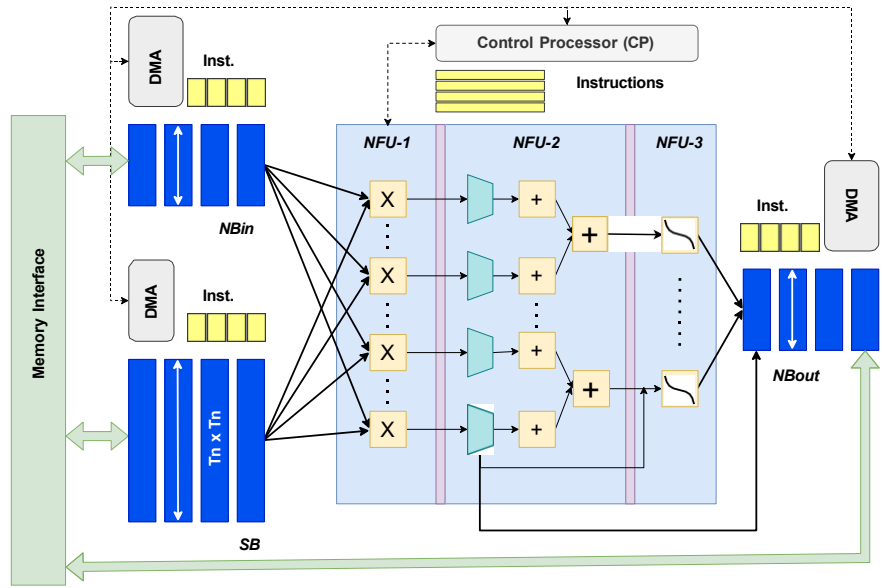

**Fig. 1.5** DaDianNao Accelerator Architecture (the figure is recreated from [20]).

- **DaDianNao** builds on the DianNao design but scales it up with multiple inter-connected tiles. This expansion allows it to handle larger neural networks more efficiently. A distinctive feature of DaDianNao is the integration of a large on-chip embedded DRAM (eDRAM). This eDRAM stores neural network parameters, significantly reducing data movement and enhancing computational speed. By having each tile work in parallel, DaDianNao achieves a higher processing capacity, making it well-suited for more complex neural network tasks. The architecture is depicted in Fig. 1.5.

- **ShiDianNao** takes a different direction, focusing on low-power applications, particularly in image processing. Its architecture is simplified compared to DianNao, aiming to reduce energy consumption. This makes ShiDianNao ideal for embedded systems where power efficiency is crucial. The design targets high-throughput vision processing tasks, optimizing the architecture for embedded applications that require efficient image processing capabilities.
- **PuDianNao** extends the capabilities of the DianNao series to support a broader range of machine learning algorithms, not just neural networks. It includes specialized hardware units dedicated to different machine-learning tasks, offering a more versatile approach to machine-learning computations. This general-purpose orientation makes PuDianNao a flexible solution, capable of handling various types of machine learning algorithms beyond the scope of traditional neural network processing.

### 1.3.3 The Neural Processing Unit (NPU)

The Neural Processing Unit (NPU) falls under the category of a digital ASIC. NPU operates by transforming select segments of general-purpose programs into neural network (NN) models [17]. This process, known as the Parrot transformation, involves identifying program segments that can tolerate approximation without significantly impacting overall accuracy. Once transformed, these segments are executed on the NPU, which is specially designed to efficiently process NN models. The integration of NPUs into traditional computing systems involves a co-processor model, where NPUs work alongside standard CPUs. The CPU handles regular precise computations, while the NPU accelerates approximate computations. This dual-processor approach allows for a balance between accuracy and computational efficiency. The architecture of the NPU, particularly in the operational framework depicted in Figure 6, showcases the detailed workflow of how program segments are processed. This includes data flow, control mechanisms, and the interaction between the NPU and the main CPU. The advantage of this setup lies in its ability to significantly improve performance and reduce energy consumption for suitable tasks, with the trade-off being a manageable decrease in computational accuracy. The Neural Processing Unit (NPU) architecture is shown in Fig. 1.6. This specialized processing unit for AI works in the following steps at the architectural level:

1. Identification of Approximable Code Segments: The CPU identifies code segments within a general-purpose program that are suitable for approximation.
2. Parrot Transformation: These identified segments are transformed into neural network models using the Parrot transformation. This involves mapping the program's logic and data flow into a neural network structure.

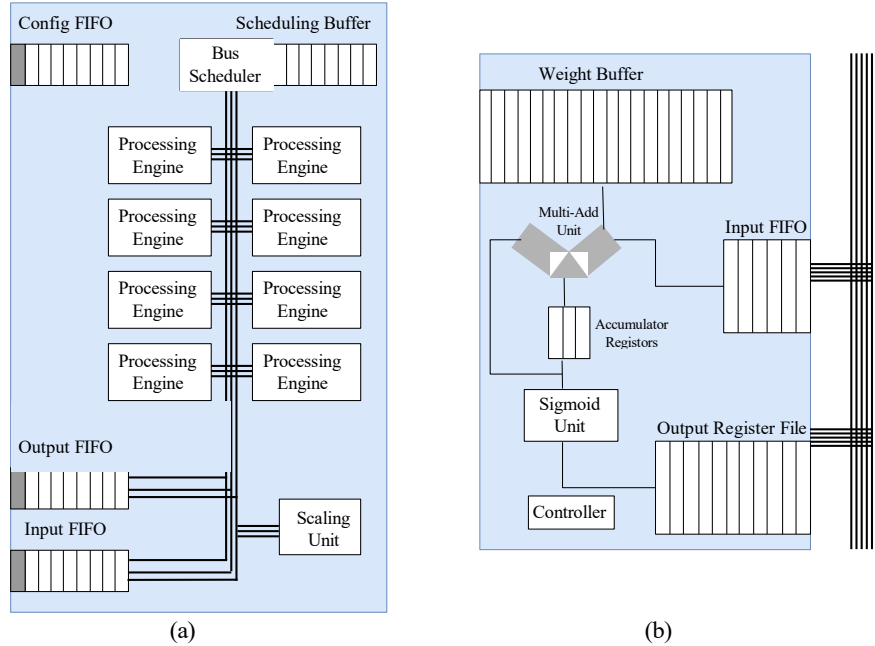

**Fig. 1.6** The Neural Processing Unit (NPU) Architecture (a) 8-Processing engine (PE) NPU, (b) Single processing engine (PE) (the figure is recreated from [17]).

3  Execution on NPU: The transformed neural network model is then executed on the NPU. The NPU is specialized for efficient neural network computations, particularly those involving operations like matrix multiplications and activations.

4  Interaction with CPU: The NPU operates as a co-processor to the CPU. It handles the approximate computations, while the CPU continues to process tasks requiring high accuracy.

5  Output Integration: The results from the NPU are integrated back into the main program flow on the CPU, ensuring that the overall program objectives are met despite the approximation in certain segments.

6  Performance and Energy Efficiency: The NPU provides enhanced performance and energy efficiency for the approximate computing tasks, with a tolerable reduction in accuracy.

This architecture leverages the strengths of both neural network processing and traditional CPU computations to achieve an optimal balance of speed, energy efficiency, and accuracy.

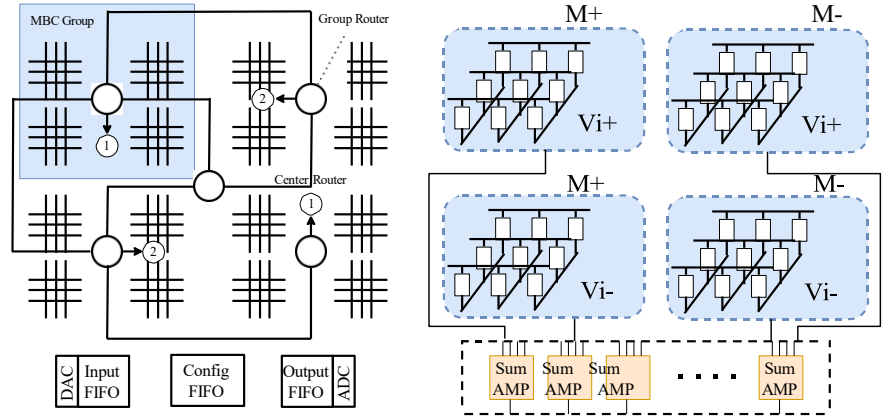

**Fig. 1.7** The RENO Architecture (The figure is recreated from [19]).

### 1.3.4 RENO Architecture

RENO is a highly efficient, reconfigurable accelerator designed for neuromorphic computing [19], which falls under the Neuromorphic IC category among different hardware solutions for AI. It leverages memristor-based crossbar arrays for enhanced mixed-signal computation, speeding up artificial neural network (ANN) executions. RENO's architecture consists of hierarchically arranged memristor-based crossbar (MBC) arrays that support various ANN topologies through a mixed-signal interconnection network (M-Net). The design demonstrates significant performance speedups and energy savings compared to general-purpose processors and other digital neural processing units. It balances efficiency, fault tolerance, and flexibility in neural network computations. The architecture is depicted in Fig. 1.7 and different features of this architecture are listed below:

1. The Core of RENO: At the heart of RENO lies its Memristor-Based Crossbar (MBC) arrays. These arrays are fundamental to RENO's ability to process artificial neural networks (ANNs) efficiently. Unlike traditional computing systems, RENO's MBC arrays enable simultaneous memory storage and computation, significantly enhancing processing speeds and reducing energy consumption.

2. Hierarchical Architecture: RENO's architecture is hierarchically structured, allowing for flexible and efficient handling of various ANN topologies. This hierarchical design not only streamlines computation but also provides a framework for scaling and integrating larger and more complex neural networks.

3. Mixed-Signal Interconnection Network (M-Net): A pivotal component of RENO is its mixed-signal interconnection network, known as M-Net. M-Net facilitates communication between different layers and elements of the neural network, ensuring seamless data flow and processing across the entire architecture.

4 Performance and Efficiency: When compared to general-purpose processors and digital neural processing units, RENO demonstrates a substantial leap in performance. It achieves remarkable speedups in ANN executions, coupled with significant energy savings. This efficiency is a result of its mixed-signal computation approach and the inherent advantages of memristor technology.

5 Balancing Flexibility and Fault Tolerance: RENO's design adeptly balances efficiency with flexibility and fault tolerance. It can adapt to various neural network structures, offering a reconfigurable platform for diverse computing needs. Additionally, RENO incorporates mechanisms to handle potential faults in memristor arrays, ensuring reliable operation.

RENO stands as a testament to the innovative advancements in neuromorphic computing. With its unique architecture and superior performance, RENO paves the way for new possibilities in AI and machine learning applications. As technology evolves, RENO's impact on neuromorphic computing will undoubtedly be a subject of keen interest and continued research.

### 1.3.5 Neurocube Architecture

Neurocube, introduced in 2016 redefines neuromorphic computing with its unique digital architecture integrated with high-density 3D memory [24]. This design is specifically aimed at enhancing neural network computations, leveraging the depth of 3D memory to achieve superior computational performance and power efficiency. The architecture is depicted in Fig. 1.8 has the following features:

1 In-Memory Neuromorphic Processing: At its core, Neurocube combines computing layers within a 3D memory stack. This allows for direct processing within memory layers, reducing the data movement significantly.

2 Memory-Centric Neural Computing (MCNC): Neurocube utilizes the MCNC approach, harnessing memory-centric techniques for efficient data-driven neural algorithms. This leads to a substantial increase in both energy and computational efficiency.

3 Programmable Neurosequence Generator: Key to its flexibility, Neurocube employs memory-based state machines, enabling it to program and adapt to a wide range of neural network algorithms.

4 Distinctive Features: Unlike conventional architectures that distinctly separate memory and computation, Neurocube's integration of a compute layer within a 3D memory stack sets it apart, resulting in higher operational efficiency and reduced latency.

Neurocube offers several advantages, including enhanced computational efficiency, where it excels in neural computations by providing rapid processing speeds with lowered energy consumption. It also boasts scalability and flexibility, matching GPU-like programmability while offering more efficient power and computational handling. Additionally, Neurocube's advanced memory integration, through the incorporation of 3D high-density memory, significantly boosts memory bandwidth and minimizes latency. However, Neurocube has limitations such as design complexity, with its advanced structure posing challenges in design and manufacturing. Moreover, its application might be limited compared to more general-purpose architectures, as it is specifically tailored for neuro-inspired algorithms.

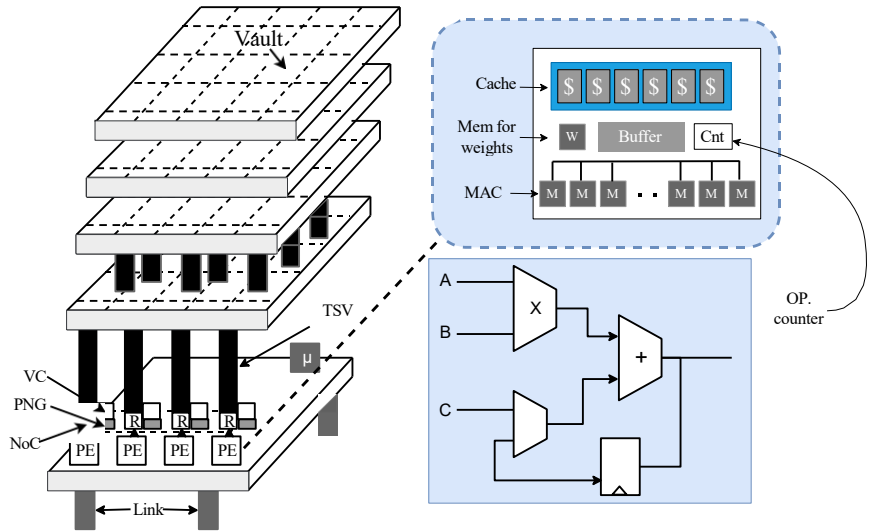

**Fig. 1.8** Neurocube Architecture (left) and Processing Elements (PEs) Organization (right) (the figure is recreated from [24]).

### 1.3.6 PRIME: ReRAM based Processing-in-memory Architecture

PRIME (Processing in ReRAM-based Main Memory) is an innovative architecture leveraging the unique capabilities of Resistive Random Access Memory (ReRAM) [28]. It's designed to enhance neural network computation efficiency by integrating processing directly into memory, addressing the "memory wall" issue prevalent in traditional computing systems. The detail architecture features are as follows:

1  ReRAM Crossbar Arrays: Central to PRIME is the ReRAM crossbar arrays, enabling matrix-vector multiplication, a critical operation in neural networks. ReRAM-based crossbar arrays perform matrix-vector multiplication differently than CMOS-based systems. Each crossbar junction represents a synaptic weight, and the array inherently executes MAC operations through Ohm's Law and Kirchhoff's Law, making the process highly parallel and energy-efficient.

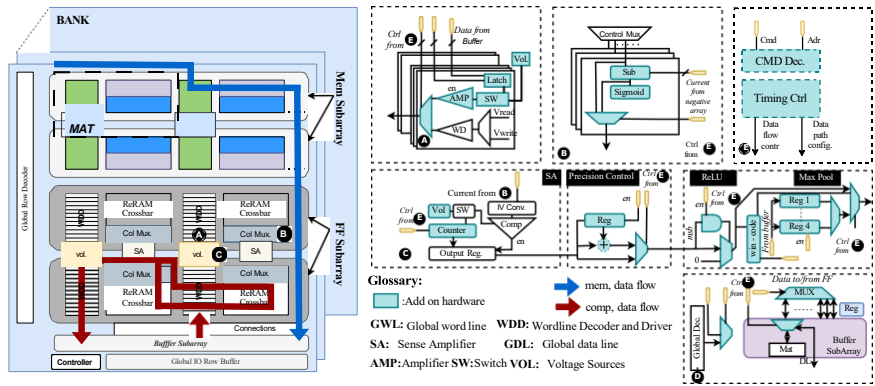

**Fig. 1.9** The PRIME Framework Illustrated. On the left, the memory bank layout is shown, with bold blue and red lines indicating pathways for standard memory operations and computational tasks. On the right, enhancements specific to PRIME are highlighted: (A) a wordline driver with multiple voltage levels; (B) a column multiplexer with analog subtraction and sigmoid functions; (C) a versatile sense amplifier with counters for multi-level signals, including ReLU functions and 4-to-1 max pooling; (D) the linkage between flip-flop (FF) and Buffer subarrays; (E) the PRIME management unit. (the figure is recreated from [28])

2  Processing-in-Memory (PIM): The architecture adopts a PIM approach, combining storage and computation in memory arrays to reduce data movement. This integration allows for simultaneous data storage and processing, significantly reducing the need for data to be moved between separate memory and processing units, which is a major source of energy consumption and latency in traditional architectures.

3  Configurable Arrays: PRIME features configurable ReRAM arrays, which can switch between functioning as neural network accelerators or as standard memory storage. This flexibility enables the system to adapt to various computational demands efficiently, providing the advantage of using the same hardware for different purposes - either for high-performance neural network computations or for regular memory storage tasks.

4  Software/Hardware Interface: PRIME includes a sophisticated software/hardware interface that ensures seamless integration and implementation of diverse neural network models. This interface acts as a bridge between the computational hardware and the software algorithms, enabling the effective execution of various neural network architectures and paradigms on the PRIME platform. This design facilitates versatility and ease of adaptation to different neural network requirements and models.

PRIME offers several advantages due to its innovative design. PRIME enhances efficiency in neural network computations through its ReRAM-based processing-in-memory (PIM) architecture, enabling simultaneous processing and storage of data within the memory arrays. This design reduces the need for time-consuming and energy-intensive data transfers between separate processing and storage units. Furthermore, the PIM approach inherently minimizes data movement, significantly cutting down on energy consumption and improving operational efficiency by processing data directly within the memory arrays. PRIME's configurable ReRAM arrays provide remarkable flexibility, supporting various neural network architectures and accommodating a wide range of computational needs, enhancing the architecture's utility across different application scenarios. However, PRIME also has several limitations, primarily due to its reliance on ReRAM technology. The integration of processing and memory in ReRAM arrays increases the complexity of design and fabrication, posing challenges in integrating ReRAM technology at a large scale while ensuring consistent performance. Additionally, ReRAM is not as standardized as CMOS technology and may face reliability issues such as retention, cycle-to-cycle, and device-to-device variations, which can affect the performance and predictability of PRIME. The architecture is specifically designed for neural network computations, limiting its effectiveness for other types of computations and restricting its utility in more general-purpose computing scenarios. Lastly, ReRAM technologies are known for variability and stability challenges, including retention time, endurance, and variability in electrical characteristics, which can lead to performance inconsistencies and impact the overall efficiency and reliability of the PRIME architecture.

### 1.3.7  Tensor Processing Unit (TPU):

The Tensor Processing Unit (TPU) is an AI ASIC developed by Google, tailored for machine learning tasks [23]. TPUs are designed to accelerate neural network computations, primarily focusing on inference operations rather than training. TPUs address the need for specialized hardware to handle the computational demands of large-scale neural networks efficiently. They offer significant speed-ups and energy efficiency over general-purpose CPUs and GPUs in machine-learning applications. The details of TPU architecture are depicted in Fig. 1.10. TPU has the following architectural features:

1 Matrix Multiply Unit: The TPU's core is a matrix multiply unit, which performs high-throughput 8-bit multiply-and-add operations, essential for NN computations.

2 Memory and Data Management: It includes large on-chip memories to reduce external memory access. A notable feature is the Unified Buffer, which stores intermediate data, and a Weight FIFO for staging weights.

3 Execution Model: TPUs follow a deterministic execution model, differing from the time-varying optimizations in CPUs and GPUs. This model helps meet the stringent latency requirements of NN applications.

4 Systolic Data Flow: The architecture utilizes a systolic array approach for data flow, enhancing energy efficiency and reducing the read/write operations from the unified buffer.

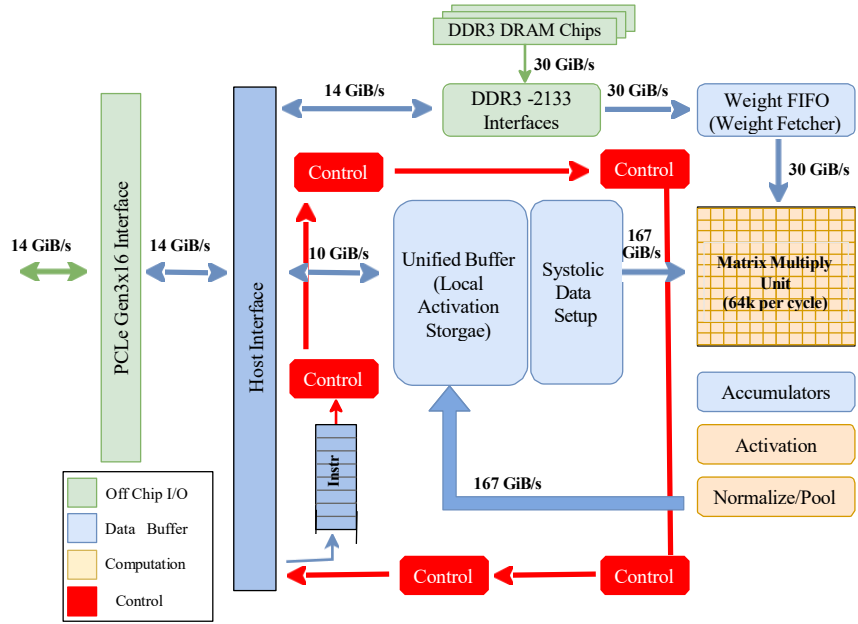

**Fig. 1.10** Tensor Processing Unit (TPU) Architecture Block Diagram (the figure is recreated from [23])

5  Software Compatibility: TPU is designed to be compatible with existing machine learning frameworks, such as TensorFlow, ensuring easy integration into existing workflows.

6  Host Integration: Operates as a coprocessor connected via the PCIe I/O bus, allowing seamless integration into existing server architectures.

7  Performance Efficiency: TPUs are optimized for high throughput per watt, significantly outperforming contemporary CPUs and GPUs in efficiency for machine learning tasks.

TPUs offer several key advantages for neural network inference tasks. They are highly efficient and deliver exceptional throughput, significantly outperforming traditional CPUs and GPUs in these applications. TPUs also consume less power compared to CPUs and GPUs, making them more energy-efficient. Additionally, TPUs are designed to handle low-precision computations effectively, which are common in neural network operations, further enhancing their performance in these specific tasks. However, TPUs also have certain limitations. They are less flexible for tasks outside of neural network inference, making them less versatile compared to general-purpose processors. TPUs also rely heavily on specific software ecosystems, such as TensorFlow, which can limit their usability with other software platforms. Moreover, TPUs come with higher upfront costs and complexity compared to general-purpose processors, which can be a barrier to adoption for some users.

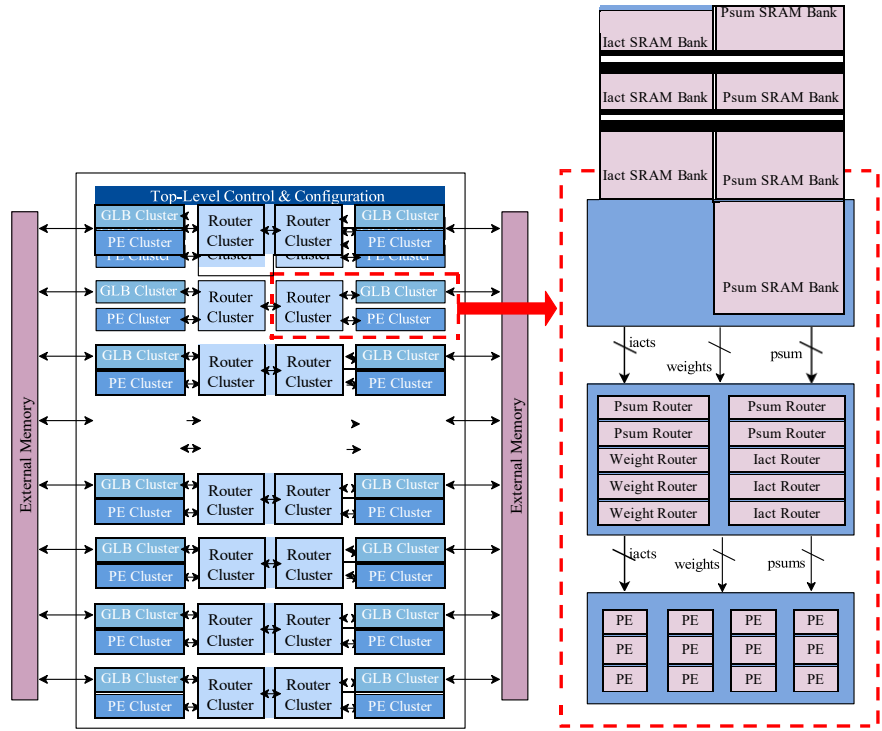

**Fig. 1.11** Eyeriss v2 Top Level Architecture (the figure is recreated from [27]).

## 1.3.8 Eyeriss v2 Architecture

Eyeriss V2, shown in Fig.1.11, is a successor to the Eyeriss architecture [26] implemented as an ASIC. It is tailored for efficient processing of compact and sparse Deep Neural Networks (DNNs) and was published in 2019 by a research group at MIT [27]. It addresses the challenges posed by varying layer shapes and sizes and the demand for energy-efficient hardware in mobile devices. The architecture details are given below:

1. Hierarchical Mesh Network (HM-NoC): A key feature of EyerissV2, HM-NoC adapts to different bandwidth and data reuse requirements. It efficiently handles the data flow across the processor, enhancing throughput and energy efficiency.
2. Processing Element (PE) Design: The architecture includes PEs tailored for processing sparse DNNs. It utilizes Compressed Sparse Column (CSC) format for data storage and movement, reducing storage costs and improving energy efficiency.
3. Row-Stationary (RS) Dataflow: EyerissV2 adopts RS dataflow, optimizing data reuse and minimizing data movement. This approach is particularly effective for the compact DNN layers found in modern mobile-oriented networks.
4. Run-Length Coding (RLC) in EyerissV2: RLC, a compression technique, is not explicitly mentioned in EyerissV2's architecture. However, the use of CSC for data compression can be seen as a related approach to handling sparse data efficiently. This methodology plays a crucial role in optimizing data storage and transfers in sparse DNNs, directly impacting the overall performance and energy efficiency of Eyeriss V2.

Eyeriss V2 offers several advantages in deep neural networks (DNNs), particularly for mobile devices. It excels in energy-efficient processing, a crucial feature for battery-powered mobile applications. The architecture is specifically designed to handle the complexities and challenges associated with compact and sparse DNNs, which are becoming increasingly common in mobile technology. Additionally, Eyeriss V2's Hierarchical Mesh Network-on-Chip (HM-NoC) provides flexible data handling, ensuring high throughput and efficiency across various DNN structures. However, Eyeriss V2 also presents certain limitations. The advanced architecture, particularly the HM-NoC, introduces significant design and implementation complexities. Furthermore, being tailored for compact and sparse DNNs, Eyeriss V2 may have limited applicability for other types of neural networks or general-purpose computing tasks.

### 1.3.9  RLC Compressed Form Architecture (CompAct)

CompAct, implemented as ASIC, represents a breakthrough in CNN accelerator architecture, focusing on efficient activation compression to reduce power consumption [25]. It is a systolic array-based accelerator, particularly tailored for convolutional neural networks (CNNs), with an emphasis on minimizing the energy costs associated with high-frequency memory accesses. The architecture is illustrated in Fig. 1.12. The RLC compressed form architecture has the following features:

1  Run-Length Coding (RLC) Integration: CompAct's core feature is its innovative use of RLC for on-chip compression. This approach is key to reducing memory access energy, a significant concern in CNN processing. RLC is a data compression technique that encodes sequences of identical data elements. In CompAct, RLC is ingeniously adapted for compressing activation data in CNNs. This adaptation ensures efficient storage and processing, reducing the overall memory bandwidth and energy consumption typically associated with handling large volumes of neural network data. By efficiently encoding repetitive sequences, CompAct's RLC reduces the storage and transfer of redundant data, significantly impacting the accelerator's performance and energy efficiency.
2  Adaptation to CNN Workloads: The architecture is specifically optimized for CNN workloads, utilizing a modified RLC scheme that aligns with row-major scheduling of activations for efficient data compression.
3  Sparse and Lossy RLC Variations: CompAct introduces Sparse-RLC for layers with high data sparsity and Lossy-RLC for more aggressive compression. These variations balance the trade-off between compression efficiency and neural network accuracy.
4  Look-Ahead Snoozing (LAS) Technique: A novel LAS technique is employed to decrease leakage energy in activation buffers, further enhancing energy efficiency.

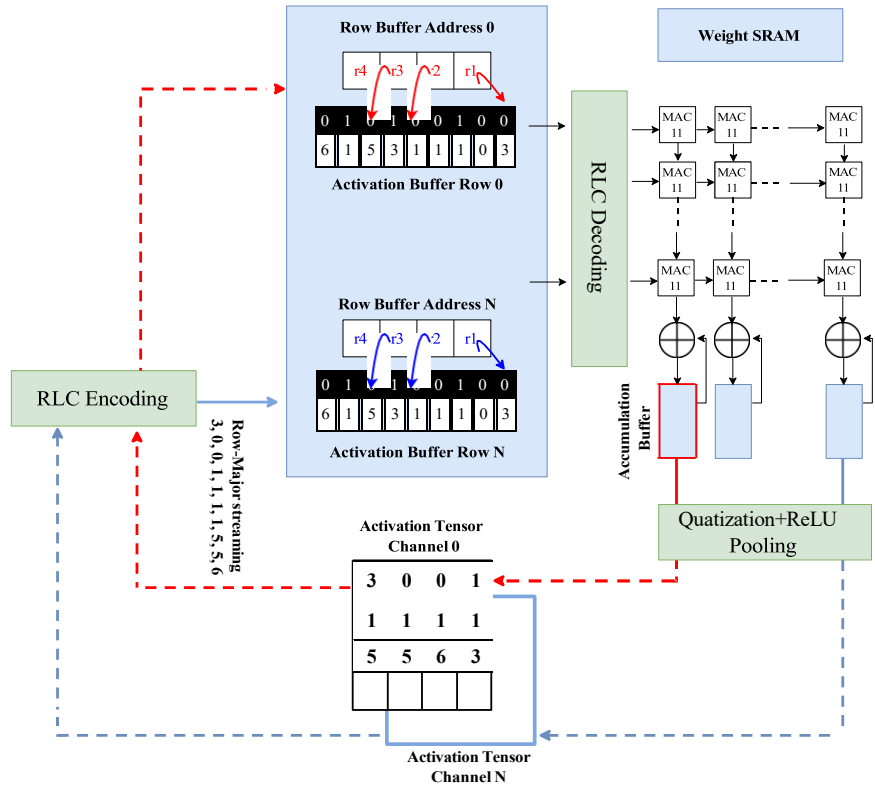

**Fig. 1.12** Illustration of the CompAct Architecture. The RLC encoding and decoding blocks ensure that activations are always stored on-chip in a compressed form(the figure is recreated from [25]).

5 Systolic Array-Based Design: The design utilizes a systolic array structure, which is advantageous for parallel processing in neural network computations.

CompAct's integration of Run-Length Coding (RLC) for on-chip compression addresses the specific challenge of memory access energy in Convolutional Neural Networks (CNNs). This architecture stands out due to its ability to switch between Sparse and Lossy RLC, offering a customizable approach to balance compression and accuracy. CompAct significantly enhances energy efficiency by reducing the activation buffer energy, leading to lower total chip energy consumption. Its tailored RLC scheme effectively minimizes the memory footprint, enhancing performance without sacrificing accuracy. Furthermore, the architecture's adaptability to different CNN workloads makes it versatile and efficient in various application scenarios. However, the sophisticated integration of RLC and its variations introduces complexity in design and implementation. Additionally, especially with Lossy-RLC, there is a potential trade-off between compression efficiency and the accuracy of neural network outputs. In summary, CompAct is a sophisticated and efficient accelerator architecture for CNNs, combining innovative compression techniques and energy-saving strategies to address the specific challenges of neural network computations.

## 1.4  Design Considerations and Optimization Technique

### 1.4.1  Design Considerations

Designing an AI accelerator involves various considerations, shaped by the unique demands of neural network computations and the specific strengths of different architectures:

- **Efficiency in Data Movement:** Efficiency in data movement is a pivotal consideration in designing neural network accelerators, as seen in the architectures of PRIME, Neurocube, EyerissV2, and TPU. These architectures demonstrate innovative approaches to minimize data transfers, a significant bottleneck in performance and energy consumption. For instance, PRIME's processing-in-memory architecture, leveraging ReRAM crossbar arrays, integrates computation within memory units, significantly reducing data movement. Similarly, the TPU's systolic array design enables efficient data flow through its matrix multiply units, minimizing the need for frequent memory access. In practical terms, AI accelerator design should focus on minimizing memory access. This involves optimizing data paths and storage within the architecture to ensure that data is processed as close to its storage location as possible, thereby reducing the energy and time costs of data movement. The goal is to create a balance between processing power and data accessibility, tailoring the architecture to handle the specific demands of neural network computations efficiently.
- **Adaptability to Network Sparsity:** Adaptability to network sparsity is a critical design consideration for neural network accelerators, as sparsity in neural networks can significantly impact computational efficiency and energy consumption. Efficient processing of sparse neural networks requires architectures that can dynamically adapt to varying levels of sparsity. This involves employing techniques such as sparse matrix operations, which can skip over zero-valued elements to reduce unnecessary computations, and incorporating dynamically configurable units that activate only for relevant computations. For instance, the architecture of EyerissV2, designed to handle compact and sparse DNNs, exemplifies this approach. It adopts strategies that optimize for sparsity, such as compressed data formats and selective processing, to improve energy efficiency and performance. Designers can learn from these architectures to create systems that are more adept at handling the unique challenges posed by sparse neural networks.
- **Energy Consumption:** Energy efficiency is paramount in neural network accelerators, especially for applications in embedded systems. Beyond quantization and dynamic voltage and frequency scaling (DVFS), other techniques like low-power memory design are crucial. This involves using energy-efficient memory types, such as SRAM or non-volatile memory, which consume less power than traditional DRAM. Clock gating and power gating techniques can also be employed to reduce power usage in idle circuitry. Additionally, optimizing the memory hierarchy to reduce access to higher power-consuming memory units, and employing approximate computing methods where high precision is not required, can further lower energy consumption. These strategies not only reduce the power usage during intensive computations but also minimize the overall energy footprint of the hardware, making the accelerators more suitable for energy-constrained environments.

- **Scalability:** Scalability in neural network accelerators is critical to adapt to the diverse requirements of different neural network models. This extends beyond simple modular design to encompass a range of techniques aimed at ensuring the accelerator can handle networks of varying sizes and complexities. Adaptive precision, where the computational precision is adjusted based on the requirement of the task, is one approach. This not only aids in resource management but also in energy efficiency. Scalability also involves employing advanced interconnect architectures that can efficiently manage the increased data flow in larger networks. These architectures should support both vertical scaling, where the capabilities of an individual unit are enhanced, and horizontal scaling, where more units are added to increase throughput. Another aspect is the use of reconfigurable computing elements, like FPGAs, which allow the accelerator to adapt to different network topologies. Additionally, the integration of heterogeneous computing elements, combining CPUs, GPUs, and custom ASICs, can provide the flexibility to optimize performance for various types of neural network computations. Lastly, scalability must also consider the software stack, ensuring that the accelerator's hardware is fully utilized through efficient software frameworks and libraries that can adapt to the changing hardware configurations.

- **Hardware-Software Synergy:** Optimal performance in neural network accelerators hinges on the deep integration between hardware and software. This goes beyond using specialized hardware instructions or adapting hardware to specific algorithms. It involves developing software that can dynamically interact with hardware features, including real-time reconfigurability and adaptability to various computational loads. Advanced machine learning techniques, such as hardware-aware neural architecture search (NAS), play a pivotal role in this co-optimization process, allowing for the simultaneous design of both hardware and software to maximize performance and efficiency. Additionally, the synergy should consider the use of compiler optimizations that can translate high-level neural network models into efficient hardware-specific instructions, bridging the gap between algorithmic design and hardware execution. This also includes the development of robust APIs and SDKs that provide an abstraction layer, enabling developers to efficiently utilize hardware resources without needing deep hardware expertise. Furthermore, embracing emerging technologies like in-memory computing and neuromorphic computing in the hardware design can open new avenues for software to exploit these novel architectures, leading to breakthroughs in processing speed and energy efficiency. This comprehensive approach ensures that the hardware is not only capable of running current neural network models efficiently but is also future-proofed against the evolving landscape of machine learning algorithms.

- **Processing Parallelism:** Handling the vast parallel computations in neural networks demands architectures capable of concurrent processing. This extends beyond mere multiple processing elements working in unison. One example is the use of systolic arrays, as seen in Google's TPU, where data flows across a grid of processors, enabling highly efficient matrix operations. Another approach is the employment of SIMD (Single Instruction, Multiple Data) architectures, commonly found in GPUs, which are adept at handling vectorized operations. Additionally, techniques like pipelining can be leveraged, where different stages of a computation are overlapped, and data parallelism, where the same operation is performed on different data sets in parallel. More advanced techniques include spatio-temporal parallelism, which combines both spatial and temporal aspects to maximize resource utilization. This approach can be particularly effective in recurrent neural network (RNN) computations, where both spatial and temporal data dependencies are prevalent. Moreover, emerging technologies like neuromorphic computing, which mimics the parallel processing capabilities of the human brain, present new frontiers in parallel processing architecture. These technologies offer the potential for massively parallel processing capabilities, far exceeding traditional computing architectures.

In summary, AI accelerator design must be evaluated for the specific needs of their applications to choose the most appropriate form of design trade-offs. By understanding and leveraging these various architectural features, designers can significantly enhance the computational efficiency of AI accelerators.

### 1.4.2 Optimization Techniques

The explosive growth of deep learning applications has fueled the demand for efficient hardware accelerators. Traditional processors struggle with the immense computational demands of deep learning algorithms, leading to high energy consumption and long latency. Hardware accelerators offer a promising solution by exploiting the inherent parallelism and data-intensive nature of these algorithms. However, designing efficient hardware accelerators for deep learning requires careful consideration of various optimization techniques. We will discuss these techniques in this section, exploring their methodologies and impact on accelerator performance.

**Architectural Optimization Techniques:**

**1  Dataflow Optimization:**

- Data Reuse: Deep learning models often exhibit data reuse patterns, where the same data is processed multiple times. Optimizing dataflow by exploiting these patterns can significantly reduce memory bandwidth requirements and improve performance. Techniques such as data reuse buffers and loop transformations can effectively manage data movement and minimize redundant transfers.

– Pipelining: Pipelining breaks down complex computations into smaller, independent stages that can be executed concurrently. This allows for efficient utilization of hardware resources and improves throughput. By carefully designing the pipeline stages and balancing their workload, accelerators can achieve high performance with minimal latency.

– Memory Hierarchy Design: Optimizing the memory hierarchy plays a crucial role in ensuring efficient data access. A well-designed memory hierarchy utilizes different memory types with varying access speeds and capacities to effectively manage the trade-off between performance and cost. Techniques like caching and prefetching can further improve data access latency and reduce memory bandwidth consumption.

**2  Computational Optimization:**

– Quantization: Reducing the precision of weights and activations from 32-bit floating-point to lower precision formats like 8-bit integers can significantly improve performance and reduce memory footprint. Quantization techniques like fixed-point arithmetic and custom quantization schemes can achieve significant performance gains while maintaining acceptable accuracy.

– Pruning: Pruning removes redundant or insignificant connections from the neural network, leading to a smaller model with minimal accuracy loss. This can significantly reduce computational complexity and memory requirements, which is particularly beneficial for resource-constrained environments.

– Approximate Computing: Approximate computing techniques trade-off some accuracy for significant gains in performance and energy efficiency. Techniques like stochastic computing and voltage scaling can be utilized to approximate computations in a controlled manner, leading to faster and more energy-efficient accelerators.

**3  Hardware Co-design:**

– Algorithmic-Hardware Co-design: This approach involves jointly optimizing the neural network architecture and the hardware design to achieve the best performance and efficiency. Techniques like exploiting sparsity in the network weights and designing custom computing units for specific operations can lead to significant performance gains.

– Near-memory computing: By placing processing units closer to the memory, data movement overhead can be significantly reduced, leading to improved energy efficiency and performance. Technologies like memristors and in-memory computing offer promising avenues for near-memory processing in deep learning accelerators.

**3  Emerging Technologies::**

– Neuromorphic computing: Neuromorphic chips mimic the architecture and function of the human brain, offering a potential path to highly efficient deep learning hardware. While still in its early stages, neuromorphic computing holds immense promise for future generations of deep learning accelerators.

– Quantum computing: Quantum computers leverage the principles of quantum mechanics to perform computations that are impossible for classical computers. While still under development, quantum computing has the potential to revolutionize deep learning by enabling significantly faster and more efficient training and inference.

**Impact of Optimization Techniques:**

Optimizing hardware accelerators for deep learning using the techniques mentioned above leads to several significant benefits:

- Increased Performance: Optimization techniques can substantially improve the performance of deep learning accelerators, leading to faster inference and training times. This allows for running complex models on resource-constrained devices and real-time applications.
- Reduced Energy Consumption: Optimization techniques can significantly reduce the energy consumption of deep learning accelerators, making them more sustainable and environmentally friendly. This is particularly crucial for battery-powered devices and large-scale data centers.
- Improved Cost-Effectiveness: By reducing hardware complexity and resource requirements, optimization techniques can lead to more cost-effective deep learning accelerators. This makes them more accessible to a wider range of users and applications.
- Enhanced Accuracy: Although some optimization techniques may involve sacrificing some accuracy, others can even lead to improved accuracy by better utilizing hardware resources and exploiting specific properties of deep learning models.

The field of hardware accelerator design for deep learning is rapidly evolving, with new optimization techniques constantly emerging. By leveraging these techniques effectively, engineers can design efficient accelerators that deliver high performance, low energy consumption, and improved cost-effectiveness. As deep learning applications continue to grow in complexity and demand, the development and optimization of hardware accelerators will play a critical role in enabling their widespread adoption across diverse domains.

## 1.5 Applications and Future

Hardware accelerators are used in various applications where deep learning models play a critical role. These applications include:

### 1.5.1 Revolutionizing Industries

The Diverse Applications of Hardware Accelerators for Deep Learning Deep learning models have revolutionized countless industries, but their complex computational needs pose significant challenges. Hardware accelerators, specialized hardware designed to efficiently execute these models, are emerging as the key to unlocking their full potential. By dramatically reducing training times and inference speeds, they enable faster innovation and deployment across diverse fields.

**1 Computer Vision:**

– Image and Video Recognition: From facial recognition in security systems to automatic image tagging on social media platforms, hardware accelerators power the real-time analysis of vast amounts of visual data.
– Object Detection: Self-driving cars rely on hardware accelerators for real-time object detection on the road, ensuring safe navigation. Similarly, retail stores utilize these tools for inventory management and theft prevention.
– Autonomous Vehicles: The development of autonomous vehicles hinges on the efficient processing of sensor data. Hardware accelerators enable real-time obstacle detection and path planning, making self-driving cars a reality.

**2 Natural Language Processing:**

– Machine Translation: Breaking down language barriers is becoming easier with hardware accelerators. They power real-time translation tools, facilitating communication across cultures and languages.
– Sentiment Analysis: Businesses rely on sentiment analysis to understand customer opinions and improve their products and services. Hardware accelerators accelerate this process, allowing for faster and more accurate insights.
– Chatbot Development: Chatbots are taking over customer service and online interactions. Hardware accelerators enable natural and engaging conversations by powering the complex algorithms behind these virtual assistants.

**3 Speech Recognition:**

– Voice Assistants: From Siri to Alexa, voice assistants are changing the way we interact with technology. Hardware accelerators power these tools, enabling natural language understanding and accurate response generation.
– Smart Speakers: Smart speakers are revolutionizing the way we listen to music and access information. Hardware accelerators make these devices responsive and intelligent, providing users with a seamless experience.
– Transcription Services: Transcribing audio and video content is becoming easier and more efficient with hardware acceleration. This technology is transforming workflows in various industries, including education, media, and law.

**4 Robotics:**

- Navigation: Robots rely on accurate navigation to operate in complex environments. Hardware accelerators enable them to perceive their surroundings and plan safe and efficient routes.
- Object Manipulation: Robotic arms are becoming increasingly sophisticated, performing delicate tasks in manufacturing and healthcare. Hardware accelerators ensure the precise and controlled movement of these robots.
- Decision-Making: Robots are being tasked with making complex decisions in autonomous systems. Hardware accelerators allow them to process information quickly and accurately, leading to better decision-making capabilities.

**5  Healthcare:**

- Medical Image Analysis: Hardware accelerators enable the rapid analysis of medical images, such as X-rays and MRIs. This leads to faster diagnosis, more accurate treatment plans, and improved patient outcomes.
- Drug Discovery: The search for new drugs is a complex and time-consuming process. Hardware accelerators can analyze vast amounts of data, leading to faster development of life-saving medications.
- Personalized Medicine: Tailoring healthcare to individual patients is becoming a reality with hardware-accelerated deep learning. This approach leads to more effective and personalized treatment plans.

**6  Finance:**

- Fraud Detection: The financial industry relies on robust fraud detection systems to protect against financial crimes. Hardware accelerators enable real-time analysis of financial transactions, preventing fraudulent activities.
- Risk Assessment: Evaluating financial risk is crucial for investors and lenders. Hardware accelerators provide faster and more accurate risk assessments, leading to better financial decisions.
- Algorithmic Trading: High-frequency trading relies on split-second decision-making. Hardware accelerators enable faster execution of trades, providing a competitive edge in the financial markets.

**7  Internet of Things (IoT):**

- Edge Computing: Processing data directly at the edge of the network is crucial for many IoT applications. Hardware accelerators enable real-time analysis of sensor data, leading to faster response times and improved decision-making.
- Sensor Data Analysis: The vast amount of data generated by IoT devices needs to be processed efficiently. Hardware accelerators enable real-time analysis of this data, unlocking valuable insights and enabling smarter devices.
- Anomaly Detection: Identifying unusual patterns in sensor data is critical for various applications, such as predictive maintenance and fault detection. Hardware accelerators power these anomaly detection algorithms, ensuring timely identification and response to potential problems.

### 1.5.2  The Future of Hardware Accelerator

The field of hardware acceleration for deep learning is experiencing a thrilling transformation, with innovative technologies and applications emerging rapidly. This exciting landscape is driven by several key trends:

1  **Orchestrating Heterogeneity:** Instead of relying on a single dominant technology, the future lies in leveraging the strengths of diverse hardware approaches. This involves combining different types of accelerators, such as GPUs, TPUs, FPGAs, and ASICs, to achieve optimal performance and energy efficiency for specific tasks.
2  **Tailoring for Unmatched Performance:** The future moves away from a one-size-fits-all approach towards domain-specific optimization. This involves designing accelerators specifically for certain deep learning models or applications through hardware-specific algorithms, model compression techniques like pruning and quantization, and hardware-aware training.
3  **Mimicking the Brain's Efficiency:** Neuromorphic computing, inspired by the brain's unparalleled processing capabilities, aims to create hardware systems that mimic its architecture. Utilizing components like memristors, these systems promise ultra-low power consumption, high parallelism, and adaptability, opening doors for future hardware acceleration advancements.
4  **Democratizing Access:** To broaden the accessibility of hardware acceleration, seamless integration with software tools and frameworks is crucial. This includes developing high-level programming interfaces, automated optimization tools, and open-source platforms to empower a wider range of developers.
5  **Collaborative Innovation:** Shaping the future of hardware acceleration necessitates a collaborative effort from diverse stakeholders. Hardware vendors must continuously refine their technologies to address evolving needs, while software developers create user-friendly tools and frameworks to unlock hardware acceleration's full potential. Researchers, meanwhile, push performance boundaries by exploring new frontiers in hardware architecture and algorithms. Finally, end-users provide invaluable feedback and drive demand for innovative solutions across various industries. By fostering this collaborative spirit and embracing emerging trends, we can unlock a future of unprecedented innovation and progress in hardware acceleration for deep learning. This will pave the way for new possibilities across diverse domains, propelling artificial intelligence to new heights.

## 1.6  Conclusion

In the chapter, a comprehensive exploration is undertaken into the evolution and significance of specialized hardware designed to optimize artificial intelligence (AI) processes. It delves into the transformative shift from conventional computing paradigms to the adoption of AI-specific hardware such as GPUs,

FPGAs, and ASICs, emphasizing their pivotal roles in augmenting the efficiency and performance of AI tasks. The chapter also encapsulates the challenges encountered in the development of these accelerators, their influence on advancing AI capabilities, and the anticipation of future innovations in this domain. By weaving together technical insights, practical examples, and forward-looking perspectives, the chapter aims to furnish a nuanced understanding of the current state and emerging trends in AI hardware development, catering to a diverse audience ranging from industry practitioners to academic researchers.